\documentstyle[twoside,fleqn,espcrc2,psfig,epsfig]{article}

\newcommand{\AmS}{{\protect\the\textfont2
  A\kern-.1667em\lower.5ex\hbox{M}\kern-.125emS}}

\title{Tests of physics beyond the Standard Model with future low energy neutrino experiments}

\author{O. G.  Miranda\address{Departamento de F\'{\i}sica, CINVESTAV-IPN, 
        A. P. 14-740 \\ M\'exico 07000 D F M\'exico}%
        \thanks{E-mail: omr@fis.cinvestav.mx
                }}

\begin{document}

\begin{abstract}
  Neutrino-electron scattering can be used to probe neutrino
  electromagnetic properties at low-threshold underground detectors
  with good recoil electron energy resolution. We study
  the sensitivity of Helium detector experiments, such as HELLAZ, for
  artificial anti-neutrino sources.
  We show that, for a $^{90}Sr-Y$ source, one expects a sensitivity 
  to the neutrino magnetic moment at the level of $\mu_{\nu}= 2 \times
  10^{-11}\mu_{B}$. 
We also report the sensitivity that
these experiments could have in
searching for an additional gauge boson in $E_6$ models.
\end{abstract}

\maketitle

\section{Introduction}

The solar neutrino problem has motivated the study of new solutions
based on neutrino physics. The most popular solutions are based on the
idea of neutrino oscillations either in vacuum or in the Sun due to
the enhancement arising from matter effects~\cite{MSW}. In addition
there is considerable interest in alternative interpretations such as
the resonant spin-flavor solution ~\cite{Akhmedov}. Recent analyisis
shows that this solution give a better fit than those obtained for
the favoured neutrino oscillation solutions~\cite{rsf,ad-hoc-profile},
altough not in a statistically significant way.

This kind of scenario has motivated the search for a neutrino magnetic
moment by using reactor experiments such as MUNU~\cite{MUNU}. In this
talk I will concentrate on a different idea~\cite{mssv}: the use of a
radioactive isotope source with a low-energy detector such as Helium
detectors (HELLAZ-HERON). These detectors are sensitive to the
antineutrino flux through  neutrino-electron scattering.

I will also discuss the proposal~\cite{msv} of using the same type of
experiments as a test of the electroweak gauge
structure. 
It will be shown that these experiments could give complementary tests 
of physics beyond the Standard Model. 

\section{Experimental prospects for neutrino magnetic moment searches}

The use of low energy experiments in order to constraint the neutrino
magnetic moment (NMM) has been widely discussed in the literature.
The stronger bound comes from a reactor experiment \cite{Derbin} and
gives $\mu_{\nu}=1.8\times 10^{-10}\mu_B$. The MUNU collaboration is
now running and tries to improve this constraint down to
$\mu_{\nu}=3\times 10^{-11}\mu_B$ by measuring a reactor antineutrino
flux with a new detector~\cite{MUNU}.

The idea of using artificial neutrino sources (ANS) to search for a
NMM was first proposed by Vogel and Engel~\cite{Vogel}. Since then,
there has been several experimental proposals going in this
direction. LAMA collaboration is planned to search for a NMM of the
order of $10^{-11}\mu_B$ \cite{Barabanov}.  BOREXINO \cite{BOREXINO}
has also been proposed as an alternative to search for a NMM
\cite{Fiorentini}. Recently this proposal has been studied taking into
account a $^{90}~Sr$ source \cite{montanino,sinev}; in this case, a
sensitivity of $\mu_{\nu}\sim 1.6\times10^{-11}\mu_B$ seems to be
reachable. A different proposal is the use of an intense ANS with a
neutrino energy of few KeV~\cite{trofimov} . In this cases a low mass
detector is needed.

Here we discuss the potential of an artificial neutrino source in
testing the NMM in a large mass detector with
both angular and recoil electron energy resolution~\cite{mssv}. I will
concentrate on the case of Helium detectors proposals such as
HELLAZ~\cite{HELLAZ}.

For fhis purpos a $^{90}Sr-^{90}Y$ anti-neutrino source is
considered. This source has been studied by a Moscow group
\cite{Bergelson} and its potential has been discussed for the BOREXINO
case \cite{montanino,sinev}.  In order to get a number for the HELLAZ
sensitivity to the neutrino magnetic moment we have made a similar
analysis to the one performed in Ref.  \cite{montanino}. There are
other experimental proposals that consider Helium~\cite{HERON} (or
Xenon~\cite{Xenon}) as a target for detecting neutrino-electron
scattering. The following analysis could be extended to study them.

The differential cross section for the process $\nu_e e\to \nu_e e$ is
given by
\begin{eqnarray}
\frac{d\sigma}{dT}  &=& 
 \frac{2m_{e}G^2_{F}}{\pi} \big\{ 
    (g_{L}+1)^2 + g_{R}^2 (1-\frac{T}{E_{\nu}})^2  - \nonumber
 \\ & &  \frac{m_e}{E_{\nu}} (g_{L}+1)
    g_{R}\frac{T}{E_{\nu}} \big\} ,  \label{DCS}
\end{eqnarray}
where $T$ is the recoil electron energy, and $E_{\nu}$ is the neutrino
energy.  In the SM case we have $g_{L,R}=\frac12 (g_{V}\pm g_{A})$,
with $g_{A}=-\rho_{\nu e}/2$ and $g_{V}=\rho_{\nu e}(-1/2+2\kappa
\mbox{sin}^2\theta_{W})$ where $\rho_{\nu e}$ and $\kappa$ describe
the radiative corrections for low-energy $\nu_{e} e\to\nu_{e} e $
scattering~\cite{Sirlin}. For the case of $\overline{\nu_{e}} e \to
\overline{\nu_{e}} e$ scattering we just need to exchange $g_{L}+1$
with $g_{R}$ and vice versa.

If neutrino has a magnetic moment $\mu_{\nu}$, there will be an
additional contribution, given as
\begin{eqnarray}
\frac{d\sigma^{mm}}{dT}  =
 \frac{\pi\alpha^2\mu^2_{\nu}}{m^2_{e}}\big\{  \frac1T -\frac{1}{E_{\nu}}
\big\},  \label{MMCS}
\end{eqnarray}
which adds incoherently to the weak cross section. 

For the case of HELLAZ an angular resolution of 35 mrad is expected
and the expected electron recoil energy resolution is
$\sigma_T/keV=22\sqrt{T/MeV}$ \cite{HELLAZ}.  Therefore, we need to
integrate the differential cross section (either electroweak or
electromagnetic) over the error funtions:
\begin{eqnarray}
\langle\sigma\rangle=\int &d&(\theta)dT
  W(\theta )W(T) \lambda(\theta
 ,T) \nonumber \\
&&\times\frac{d\sigma^{W}}{dT}\frac{m_{e}pT}{(pcos\theta -T)^2}
\end{eqnarray}
with
\begin{equation}
W(T)=\frac{1}{2}
\left[{\rm Erf}\left(\frac{T_{2}-T}{\sqrt{2}\sigma_{T}}\right)-
      {\rm Erf}\left(\frac{T_{1}-T}{\sqrt{2}\sigma_{T}}\right)\right]
\end{equation}
and $W(\theta)$ a similar expresion for the electron recoil angle.
Here $T_1=100$~KeV $T_2=1$~MeV and $\theta_1=0$ and
$\theta_2=\mbox{arccos}(\frac{T}{\sqrt{T^2+2m_eT}})$. $\lambda(\theta
 ,T) $ accounts for the antineutrino energy spectrum. 

The total number of events will be given by
\begin{equation}
N_{0}=N_{e}\langle\sigma\rangle \times F(R,D,L)\times
\int_{t_{\rm tr}}^{t_{\rm ex}+t_{\rm tr}} dt'\, I(t'), \label{n0}
\end{equation}
where $N_e=2\times10^{30}$ is the total number of electrons in the
detector, $t_{\rm tr}=5 $~days is the source tranportation time and
$t_{\rm ex}$ is the exposure time, that we consider as 180 days,
$I(t)=I_{0}\exp(-t/\tau)$ is the intensity of the source. We are
considering $I_{0}=5$~MCi and $\tau=28$~y. The factor $F(R,D,L)$
accounts for the real fraction of the detector fiducial volume that is
sensitive to the neutrino flux and depends on the topology of the
experimental set up.  We consider the detector as a cilinder 20m long
($L=20$~m) and 5m radius ($R=5$~m) \cite{HELLAZ2}.  The fiducial
volume for this configuration will depend on the distance, $D$ from
source to the center of the detector.
In the case of a source located just at the walls of the
detector (R=D=5m) We have
$
F(R,R,L) = 0.774.
$,
For a  source located at 15 m  we will have 
$F(R,D,L)=0.919$. 

The expected background for 180 days will be,
$N_B=1980$ \cite{HELLAZ2}, and  the total $1\sigma$ uncertainty will be 
%
$
\delta_{N_{0}} =\sqrt{N_{B}+N_{0}\left(1+\delta_{A}^{2}N_{0}\right)}
$
%
with $\delta_{A}=0.01$ the stimated uncertainty of the anti-neutrino flux. 

We can compute the total number of events expected both in the
 Standard Model, as well as in the case of a non-zero neutrino
 magnetic moment:
\begin{equation}
N(\mu_{\nu})=N_0[1+\mu^2_{\nu}\frac{<\sigma^{em}>}{<\sigma^{SM}>}]
\end{equation}
Where $N_0$ es the Standard Model expectation for the number of events
computed from Eq. (\ref{n0}) and $<\sigma^{SM}>$ is the averaged
differential cross sections for the Standard Model as given by Eq. (3). 
$\mu^2_{\nu}<\sigma^{em}>$ is a similar expresion for the case of a
neutrino magnetic moment, $\mu_{\nu}$.

If the experiment measures a number of events in complete agreement with 
the SM, then we will get a bound 
\begin{equation}
\mu_{\nu}\leq\sqrt{\frac{\epsilon_{90}<\sigma^{SM}>}{<\sigma^{em}>}}
\end{equation}
with 
$
\epsilon_{90}=1.64\delta_{N_0}/N_0
$.

We can take this value as the characteristic sensitivity to a neutrino
magnetic moment search for HELLAZ.  For a source located at 5 m from the
center of the detector, we found that the sensitivity will be
$\mu_{\nu}=1.6\times10^{-11} \mu_B$. While for the more pesimistic
case of a 15 m distance the sensitivity reduces to 
$\mu_{\nu}=3.1\times10^{-11} \mu_B$. This result is similar to the 
one expected in the BOREXINO proposal
\cite{montanino}.  Although in the HELLAZ proposal a lower recoil energy
thresold is expected (100 KeV vs 250 KeV in BOREXINO) the difference
in mass (and therefore in the number of electrons) makes decrease the 
number of events. 

\section{Experimental prospects for $Z'$ searches}

The values of the coupling constants governing
$\nu_e e\to \nu_e e$ scattering in the SM have been well measured
through the $e^{+}e^{-} \to l^{+}l^{-}$ process at
LEP.  These results have given strong constraints on
the mixing of the standard Z boson with an additional
\mbox{$Z^\prime$}, in the framework of global fits of the electroweak
data~\cite{leike}. In what follos,  we will, focus 
on the possibility of probing the \mbox{$Z^\prime$ } mass at
low-energy $\nu_e e\to \nu_e e$ scattering experiments. For
convenience we define the parameter
$\gamma=\frac{M^2_{Z}}{M^2_{Z^\prime}}$ and neglect the mixing angle
$\theta^\prime$ between the SM boson and the extra neutral gauge
boson.

For extended models, the neutral current contribution to the
differential cross section will be, for $\theta^\prime=0$,
\begin{eqnarray}
&\delta&\frac{d\sigma}{dT}=\gamma\Delta = \gamma\frac{2m_{e}G^2_{F}}
{\pi}\times
\nonumber \\
&& \big\{ 
    D  + E \frac{T}{E_{\nu}}(\frac{T}{E_{\nu}}-2) - 
    F \frac{m_e}{E_{\nu}} \frac{T}{E_{\nu}}\big\} \label{dc}
\end{eqnarray}
with $\Delta$ in obvious notation and 
\begin{eqnarray}
D&=&2(g_{L}+1)\delta g_{L}+ 2g_{R} \delta g_{R} \\
E&=&2g_{R}\delta g_{R} \\
F&=& (g_{L}+1)\delta g_{R}+
    g_{R}\delta g_{L}  \label{coe}
\end{eqnarray}
where $g_{L}$ and $g_{R}$ are the SM model expressions and $\delta
g_{L,R}$ give the corrections due to new physics. The specific form of
$\delta g_{L}$ and $\delta g_{R}$ can be found in \cite{lama} for the case
 of a LRSM \cite{LR1}, and also for the case of $E_6$ models
\cite{npb345}.

The correction to the $\nu_{e} e$ scattering depends on the model. 
In the  analysis done in Ref. \cite{lama},  we showed 
that the sensitivity is bigger at $\mbox{cos}\beta
\simeq 0.8$ and it is almost zero for $\mbox{cos}\beta\simeq -0.4$. Of
the most popular models ($\chi$, $\eta$ and $\psi$ models) we can say
that the $\chi$ model is the most sensitive to this scattering.  A
similar result can be obtained for the case of anti-neutrino sources,
such as $^{147}$~Pm \cite{Kornoukhov}, proposed for the LAMA experiment
\cite{lama,Barabanov}.  
This source produces antineutrinos through the $^{147}Pm \to
^{147}Sm+e+ \overline{\nu_{e}}$ beta decay. In this case we have an
antineutrino spectrum with energies up to 235 KeV. 

The possibility of surrounding this ANS with a NaI(Tl) detector is now
under consideration by the LAMA collaboration~\cite{NUMAG}. As a first step they
plan to use a 400 tones detector (approximately $2\times 10^{29}$
electrons) that will measure the electron recoil energy from 2 - 30
KeV; the source activity will be 5 MCi. A second stage with a one tone
detector and 15 MCi of $^{147}Pm$ is under study.

We can estimate the event rates expected both in the Standard
Model as well as in extended models for the configuration discussed
a\-bo\-ve.
We can compute the expected number of events per bin in
the Standard Model. For definiteness we have considered 2 KeV width
bins.  For the case of an extra neutral gauge boson, we would expect
an excess in the number of events per bin.

In order to estimate the LAMA sensitivity to the mass of a $Z^\prime$
in the $\chi$ model we have considered an experimental set up with 5
MCi source and a one tone detector. Assuming that the detector will
measure exactly the SM prediction and taking into account only the
statistical error, we obtain a sensitivity of the order of 600 GeV at
95 \% C. L., comparable to the present Tevatron result. A more
detailed analysis can be found in ref.~\cite{lama}.

Coming back to the Helium detectors. We have considered the
experimental set up of the HELLAZ proposal, and a 5 Mci $^{51}Cr$
source. A similar source has been used to calibrate both GALLEX and
SAGE solar neutrino experiments \cite{GALLEX}.
 In this case the sensitivity for the mass of an extra gauge
boson will be $M_{Z'}\simeq 450$~GeV, if the source is located at 5 m from
the center of the detector. And $M_{Z'}\simeq260$~GeV if the distance is 15
m. For this analysis 10 KeV width bins were considered~\cite{msv}.

Finally In the case of the BOREXINO proposal. Considering the
$^{51}Cr$ neutrino source as mentioned in Ref.~\cite{montanino} we
found that the sensitivity to the $Z^\prime$ mass in the $\chi$ model
will be 305 GeV, if only the statistical error is considered. If we take into
account the background the sensitivity will decrease
to 230 GeV.

\section{Conclusions}

As a conclusion we can say that the new generation of low-energy solar
neutrino-type detectors using strong artificial neutrino sources could
give complementary information about non-standard neutrino
electromagnetic properties as well as for the structure of the
electroweak interaction.

\vspace{8pt}

\noindent {\bf Acknowledgements} 


This work was supported by CONACYT M\'exico under grant J32220-E.

I would like to thanks Javier Segura, Victor B. Semikoz, Jos\'e
W. F. Valle and the people from the LAMA collaboration, with whom the
results of this work have been done.

\end{document}